\documentclass[aps,prl,twocolumn,superscriptaddress]{revtex4-2}
\usepackage{graphicx}
\usepackage{amsmath}
\usepackage{amssymb}
\usepackage{color}
\usepackage{soul}
\definecolor{blueprl}{rgb}{0.157, 0.173, 0.569} 
\definecolor{brickred}{rgb}{0.79, 0.25, 0.33} 

\newcommand{\SIQSE}{\affiliation{1}{Shenzhen Institute for Quantum Science and Engineering, Southern University of Science and Technology, Shenzhen, Guangdong, China}}
\newcommand{\DPHY}{\affiliation{2}{Department of Physics, Southern University of Science and Technology, Shenzhen, Guangdong, China}}
\newcommand{\IQA}{\affiliation{3}{International Quantum Academy, Shenzhen, Guangdong, China}}
\newcommand{\GDKL}{\affiliation{4}{Guangdong Provincial Key Laboratory of Quantum Science and Engineering, Southern University of Science and Technology, Shenzhen, Guangdong, China}}
\newcommand{\HFNL}{\affiliation{5}{
Shenzhen Branch, Hefei National Laboratory, Shenzhen 518048, China}}

\begin{document}
\nocite{*}


\title{Coupler-Assisted Leakage Reduction for\\Scalable Quantum Error Correction with Superconducting Qubits}

\author{Xiaohan Yang}
\thanks{These authors contributed equally to this work.}
\affiliation{\SIQSE}\affiliation{\IQA}\affiliation{\GDKL}\affiliation{\DPHY}

\author{Ji Chu}
\thanks{These authors contributed equally to this work}
\email{jichu@iqasz.cn}
\affiliation{\SIQSE}\affiliation{\IQA}\affiliation{\GDKL}

\author{Zechen Guo}
\affiliation{\SIQSE}\affiliation{\IQA}\affiliation{\GDKL}

\author{Wenhui Huang}
\affiliation{\SIQSE}\affiliation{\IQA}\affiliation{\GDKL}

\author{Yongqi Liang}
\affiliation{\SIQSE}\affiliation{\IQA}\affiliation{\GDKL}

\author{Jiawei Liu}
\affiliation{\SIQSE}\affiliation{\IQA}\affiliation{\GDKL}

\author{Jiawei Qiu}
\affiliation{\SIQSE}\affiliation{\IQA}\affiliation{\GDKL}

\author{Xuandong Sun}
\affiliation{\SIQSE}\affiliation{\IQA}\affiliation{\GDKL}\affiliation{\DPHY}

\author{Ziyu Tao}
\affiliation{\SIQSE}\affiliation{\IQA}\affiliation{\GDKL}\affiliation{\DPHY}

\author{Jiawei Zhang}
\affiliation{\SIQSE}\affiliation{\IQA}\affiliation{\GDKL}

\author{Jiajian Zhang}
\affiliation{\SIQSE}\affiliation{\IQA}\affiliation{\GDKL}\affiliation{\DPHY}

\author{Libo Zhang}
\affiliation{\SIQSE}\affiliation{\IQA}\affiliation{\GDKL}

\author{Yuxuan Zhou}
\affiliation{\SIQSE}\affiliation{\IQA}\affiliation{\GDKL}\affiliation{\DPHY}

\author{Weijie Guo}
\affiliation{\IQA}

\author{Ling Hu}
\affiliation{\SIQSE}\affiliation{\IQA}\affiliation{\GDKL}

\author{Ji Jiang}
\affiliation{\SIQSE}\affiliation{\IQA}\affiliation{\GDKL}

\author{Yang Liu}
\affiliation{\IQA}

\author{Xiayu Linpeng}
\affiliation{\IQA}

\author{Tingyong Chen}
\affiliation{\SIQSE}\affiliation{\IQA}\affiliation{\GDKL}
\author{Yuanzhen Chen}
\affiliation{\SIQSE}\affiliation{\IQA}\affiliation{\GDKL}\affiliation{\DPHY}

\author{Jingjing Niu}
\affiliation{\IQA}\affiliation{\HFNL}
\author{Song Liu}
\affiliation{\SIQSE}\affiliation{\IQA}\affiliation{\GDKL}\affiliation{\HFNL}
\author{Youpeng Zhong}
\email{zhongyp@sustech.edu.cn}
\affiliation{\SIQSE}\affiliation{\IQA}\affiliation{\GDKL}\affiliation{\HFNL}

\author{Dapeng Yu}
\affiliation{\SIQSE}\affiliation{\IQA}\affiliation{\GDKL}\affiliation{\DPHY}\affiliation{\HFNL}


\date{\today}

\begin{abstract}
Superconducting qubits are a promising platform for building fault-tolerant quantum computers,
with recent achievement showing the suppression of logical error with increasing code size.
However, leakage into non-computational states, a common issue in practical quantum systems including superconducting circuits, introduces correlated errors that undermine QEC scalability. 
Here, we propose and demonstrate a leakage reduction scheme utilizing tunable couplers, a widely adopted ingredient in large-scale superconducting quantum processors.
Leveraging the strong frequency-tunability of the couplers and stray interaction between the couplers and readout resonators, we eliminate state leakage on the couplers, thus suppressing space-correlated errors caused by population propagation among the couplers.
Assisted by the couplers, we further reduce leakage to higher qubit levels with high efficiency (98.1$\%$) and low error rate on the computational subspace (0.58$\%$), suppressing time-correlated errors during QEC cycles.
The performance of our scheme demonstrates its potential as an indispensable building block for scalable QEC with superconducting qubits.
\end{abstract}


\maketitle

By redundantly encoding the quantum information of a logical 
qubit into a large Hilbert space spanned by multiple physical qubits, quantum error correction (QEC) offers a promising path to bridge the gap between the physical error rates achievable by quantum computing devices and the low logical error rates necessary for practical quantum algorithms~\cite{bravyi1998quantum,terhal2015quantum,preskill2018beyondNISQ}.
The scalability of QEC relies on the assumption that physical errors to be suppressed are sufficiently uncorrelated in both space and time~\cite{fowler2013coping,ghosh2013understanding}.
However, practical quantum systems used to create qubits possess multiple energy levels, and many quantum operations rely on non-computational energy levels~\cite{dicarlo2009demonstration,barends2014superconducting,rol2019fast,schafer2018fast,xue2022quantum,evered2023high,bluvstein2024logical}.
Leakage to these non-computational energy levels during quantum operations, even having low probability per operation, can lead to correlated errors that degrade the exponential suppression
of the logical error with scale~\cite{google2021exponential}.
To address leakage on qubits while preserving the stored information, leakage reduction units (LRUs) that are compatible with QEC circuits have been proposed~\cite{aliferis2005fault,suchara2015leakage}, with several recent experimental demonstrations on superconducting qubits~\cite{miao2023overcoming,marques2023all,lacroix2023fast,huber2024parametric}.



Surface code is one of the most prominent quantum error correcting codes due to its exceptionally high tolerance to errors and hardware-friendly nearest-neighbor connectivity~\cite{fowler2012surface}.
Superconducting transmon qubits are a particularly appealing platform for implementing surface code~\cite{reed2012realization,chow2014implementing,kelly2015state,riste2015detecting,takita2016demonstration,takita2017experimental,andersen2019entanglement,andersen2020repeated,google2021exponential,krinner2022realizing,zhao2022realization,sivak2023RealtimeQuantumError,ni2023BeatingBreakevenPoint},
where the suppression of logical error with increasing code size has been demonstrated recently~\cite{google2023suppressing}.
Enabling high-fidelity two-qubit gates and suppressing spurious interactions, high-coherence tunable couplers have played a crucial role in this milestone achievement~\cite{yan2018tunable,google2023suppressing}. 
However, the introduction of couplers also opens up new leakage channels. Long-lived leakage population on the couplers, predominantly arising from coupler-assisted two-qubit gates~\cite{foxen2020demonstrating,collodo2020implementation,xu2020high,sung2021realization}, can cause time-correlated errors by degrading the performance of subsequent gates. Furthermore, the undesired population on the couplers can propagate to adjacent couplers and induce space-correlated errors, an issue that has not yet been addressed so far~\cite{aliferis2005fault,suchara2015leakage,ghosh2015leakage,brown2019handling,varbanov2020leakage,hayes2020eliminating,bultink2020protecting,battistel2021hardware,miao2023overcoming,marques2023all,lacroix2023fast,huber2024parametric}.
A comprehensive leakage reduction scheme for superconducting processors equipped with tunable couplers is thus highly desirable.

\begin{figure}
    \centering
    \includegraphics{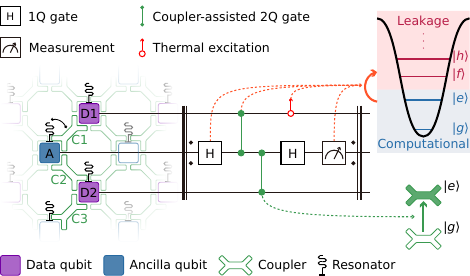}
    \caption{\label{fig1} 
    Schematic of the quantum processor and a representative repeated stabilizer measurement circuit. 
    Two data qubits D1 and D2, one ancilla qubit A, their adjacent couplers C1, C2, C3 and readout resonators are used in this experiment.
    During the stabilizer measurement, leakage to higher energy levels of qubits (red dashed) and undesired population of couplers (green dashed) can be caused by single-qubit (1Q) gates, coupler-assisted two-qubit (2Q) gates, thermal excitations and measurement, potentially leading to correlated errors.
    The leakage reduction scheme here leverages the strong frequency-tunability of the couplers and the stray interaction between the couplers and readout resonators (as indicated by the double-headed arrow).
    }
\end{figure}

In this Letter, we propose and demonstrate a coupler-assisted leakage reduction scheme leveraging the strong frequency-tunability of the couplers and the stray interaction between the couplers and readout resonators. Our scheme includes leakage reduction operations for both the couplers, designated as $c$-LRU, and the higher energy states of the transmon qubits, specifically the second and third excited states, referred to as $f$-LRU and $h$-LRU, respectively.
The $c$-LRUs are realized by simultaneously tuning all relevant couplers into a dissipative regime which is close into resonance with readout resonators. Potential space-correlated errors caused by leakage propagation among stray-coupled couplers are mitigated by the $c$-LRUs.
The $f$-LRU and $h$-LRU are achieved through state-selective qubit-coupler (QC) swaps, which are enabled by parametric frequency modulation of the coupler~\cite{caldwell2018parametrically}.
As a demonstration in the context of surface code QEC, we insert the $f$-LRUs and $c$-LRUs into a weight-two Z-stabilizer measurement circuit to return leakage population back to the computational subspace and to suppress time-correlated errors induced by artificially injected leakage.
Our demonstrations show that this scheme is capable of removing leakage on both the couplers and qubits, suppressing correlated errors in both space and time.

We demonstrate the leakage reduction scheme on a quantum processor consisting of a 2D grid of 66 transmon qubits~\cite{supplement}, with each qubit tunably coupled to its four closest neighbors--a configuration tailored precisely for the surface code. 
Physical qubits are divided into two types in surface code: data qubits for storing logical information and ancilla qubits for stabilizer measurements.
Figure~\ref{fig1} shows the layout schematic of the components on the processor used in this experiment, including two data qubits D1 and D2, one ancilla qubit A, as well as adjacent couplers C1, C2, C3 and the corresponding readout resonators. 
The resonators are coupled to the outside 50~$\Omega$ environment through a broadband Purcell filter~\cite{jeffrey2014fast}.
There are two mechanisms that give rise to the stray interaction between a coupler and the adjacent readout resonator: direct capacitive coupling and indirect coupling mediated by the qubit, yielding a net coupling strength of $\sim$25 MHz.
More details about the device performance and experimental setup are available in the Supplementary Material~\cite{supplement}.
A representative stabilizer measurement circuit for surface code QEC has also been shown in Fig.~\ref{fig1}, where different causes of state leakage are listed~\cite{motzoi2009simple,chen2016measuring,barends2019diabatic,foxen2020demonstrating,collodo2020implementation,xu2020high,negirneac2021high,sung2021realization,chen2016measuring,sank2016measurement,khezri2023measurement}.


\begin{figure*}[t]
    \centering
    \includegraphics{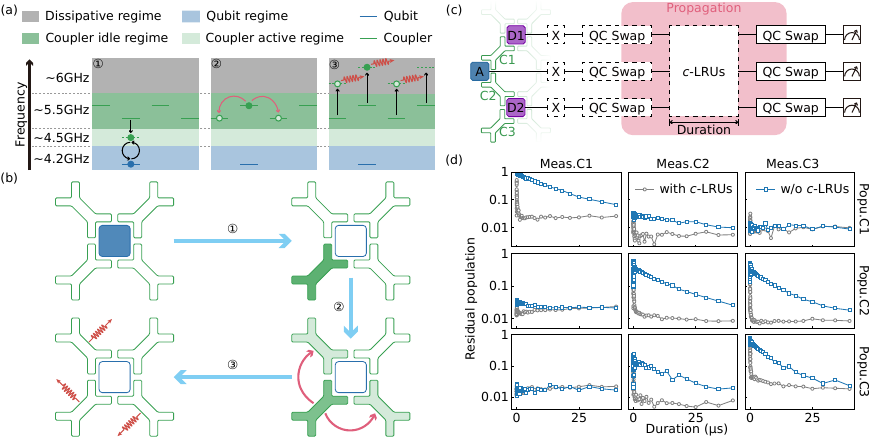}
    \caption{\label{fig2} Leakage population propagation and reduction on couplers. 
    (a)-(b) Population transfer from a qubit to its adjacent coupler through parametric QC swap ($\textcircled{1}$); Population propagation to adjacent couplers due to stray coupling ($\textcircled{2}$); and the application of simultaneous $c$-LRUs ($\textcircled{3}$). 
    The intensity of the shading in (b) indicates the amount of population on the qubit (blue) and couplers (green). 
    (c) Circuit diagram for measuring leakage propagation among couplers. C1 is populated by applying an $X$ gate on D1, succeeded by a QC swap between them. To ascertain the population on C1-C3, additional QC swaps are performed to transfer the population back to the corresponding qubits before readout.
    The three couplers are populated one-by-one by moving the beginning X gate and QC swap to their corresponding qubit respectively.
    (d) Population on the couplers with (gray) or without (blue) applying $c$-LRUs (step $\textcircled{3}$) to the couplers. 
    The three rows correspond to C1, C2 and C3 being populated respectively, whereas the three columns correspond to each coupler being measured.
    }
\end{figure*}

Figure \ref{fig2} illustrates the leakage population propagation among stray-coupled couplers and its reduction by $c$-LRUs.
We categorize the frequency spectrum into four distinct regimes, as shown in Fig.~\ref{fig2}(a): the qubit regime with an average frequency of approximately 4.2 GHz; the coupler idle regime around 5.5 GHz, where couplers reside to deactivate coupling between nearest-neighbor qubits; the coupler active regime around 4.5 GHz, where the couplers are turned into to perform two-qubit gates; and the dissipative regime around 6 GHz, where the readout resonators' frequencies are located. 
For illustrative purposes, we populate a coupler by performing a parametrically activated QC swap between an excited adjacent qubit and the coupler ($\textcircled{1}$ in Fig.~\ref{fig2}(a)-(b)). The leakage population on one coupler can inadvertently be conveyed to neighboring couplers due to stray coupling between them, particularly when its frequency path intersects with another's ($\textcircled{2}$). Electromagnetic simulations indicate that adjacent couplers in our processor exhibit a stray interaction strength of about 17 MHz~\cite{supplement}. 
Despite being relatively unexplored, the stray interaction-induced leakage propagation can result in space-correlated errors. To suppress this kind of space-correlated errors, we perform simultaneous $c$-LRUs by tuning all relevant couplers into the dissipative regime ($\textcircled{3}$), which enhances the dissipation rates due to the strong coupling with the resonators. Figure~\ref{fig2}(b) depicts the population changes during the above processes. 
The corresponding experimental sequence is depicted in Fig.~\ref{fig2}(c), where we measure the population of each coupler with an additional QC swap to transfer the population back to the adjacent qubit before readout. Figure~\ref{fig2}(d) presents the measured occupation rates for three relevant couplers, with each row representing a different coupler being populated. The population propagation between C1-C2 and C2-C3 are obvious, while C1 and C3 hardly affect each other. Without $c$-LRUs, the population on the couplers can last for tens of $\mu$s. Applying simultaneous $c$-LRUs to all couplers can rapidly dissipate the population within a timescale of hundreds of nanoseconds.
To accelerate the $c$-LRUs, rather than holding the couplers in resonance with the resonators for some duration, we can also perform rapid ($\sim 9$~ns) iSWAP operations between the couplers and resonators, subsequently allowing the photons in the resonators to decay more quickly~\cite{supplement}. We adopt this scheme in the following experiments.
To avoid undesired interference of other couplers, we have tuned all the other couplers on the processor to their sweet spot at around 7.2 GHz during this measurement, sufficiently detuned from the three couplers.

\begin{figure}
    \centering
    \includegraphics{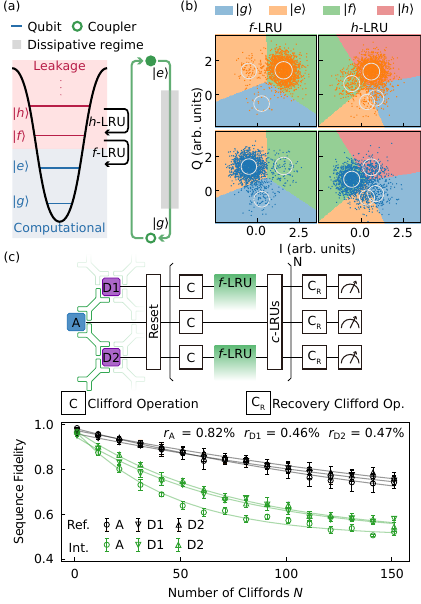}
    \caption{\label{fig3} 
    Coupler-assisted leakage reduction for the qubits.
    (a) Schematic diagram of the $f$-LRU and $h$-LRU processes. The $f$-LRU ($h$-LRU) succeeded by $c$-LRUs leads to a decrease in the qubit's total excitations by one. The coupler functions as a conduit, transferring the undesired leakage excitation from the qubit to the environment.
    (b) Single-shot readout data of qubit A for $|f\rangle$ (left), $|h\rangle$ (right), with (bottom) or without (top) applying $f$-LRU and $h$-LRU. The efficiencies for $f$-LRU, and $h$-LRU are 0.981(6) and 0.95(1) respectively. The dots in each subplot correspond to the initial 2000 measurements out of a total of $2^{14}$ shots. The white circles represent a standard deviation of the Gaussian distribution for the corresponding states.
    (c) Randomized benchmarking results with (green) or without (black) interleaved parallel $f$-LRUs on data qubits, along with simultaneous $c$-LRUs. The 'Ref.' and 'Int.' stand for 'Reference' and 'Interleaved' respectively.
    The error rates for qubit A, D1 and D2 are $r_{\rm A}=0.82\%$, $r_{\rm D1}=0.47\%$, and $r_{D2}=0.46\%$ respectively.
    On top is the circuit diagram. 
    The total duration of the $f$-LRUs and $c$-LRUs in each cycle is about 150~ns.
    A coupler-assisted reset process is applied at the beginning of the experiment.
    }
\end{figure}

With the leakage on the couplers being removed by $c$-LRUs, we can further use the coupler to assist leakage reduction of the qubit by projecting any leakage in the $|f\rangle$ or $|h\rangle$ states to $|e\rangle$ through state-selective QC swaps, which are enabled by parametric frequency modulation of the coupler~\cite{caldwell2018parametrically}. The leakage error is thus converted to Pauli error within the computational subspace and can be subsequenctly corrected by surface code. To be specific, $f$-LRU removes leakage on the qubit $|f\rangle$ state by parametrically activating a QC swap from $|fg\rangle$ (denoted as $|\rm qubit, coupler\rangle$) to $|ee\rangle$. Similarly, $h$-LRU removes leakage on the qubit $|h\rangle$ state by parametrically activating a QC swap  from $|hg\rangle$ to $|fe\rangle$, followed by the aforementioned leakage reduction on $|f\rangle$.
The added excitations on the couplers are then removed by applying $c$-LRUs simultaneously, as depicted in Fig.~\ref{fig3}(a), where couplers act as conduits that facilitate the transfer of the leakage population from the qubits to the environment.
We note that similar process can be applied to reset the qubit $|e\rangle$ state as well, increasing the rate at which experiments are carried out by initializing the qubit to its ground state before the start of the next experiment~\cite{supplement}.
For simplicity, the following discussions regarding $f$-LRU or $h$-LRU include the succeeded $c$-LRUs unless otherwise specified.
We evaluate the efficiencies of the $f$-LRU, $h$-LRU by preparing the respective state in qubit A and conducting single-shot dispersive readouts. The demodulated data in the I-Q plane, obtained with (and without) the LRU operations are depicted at the bottom (top) of Fig.~\ref{fig3}(b). 
For instance, the efficiency of $f$-LRU is quantified by
\begin{equation}
    \eta_f = 1 - \frac{ P_{ff,\rm LRU} - P_{fe} }{ P_{ff} }.
\end{equation}
Here, $P_{fe}$ denotes the measured population of $|f\rangle$ with the qubit prepared in $|e\rangle$, which may arise from readout-induced excitation~\cite{sank2016measurement}, and $P_{ff,\rm LRU},$ represents the measured population of $|f\rangle$ with the qubit prepared in $|f\rangle$ and $f$-LRU applied.
The efficiency definition of $h$-LRU is given in Ref.~\cite{supplement}.
We estimate the population of various states based on their centroids and variances, which are predetermined using a Gaussian mixture model~\cite{supplement}. To mitigate the separation error attributable to low signal-to-noise ratio (SNR), we exclude data points that fall outside one standard deviation of the Gaussian distribution. 
The efficiencies for $f$-LRU, and $h$-LRU are 0.981(6), and 0.95(1) respectively. The $h$-LRU exhibits a slightly lower efficiency than $f$-LRU, which may be attributed to the use of the second harmonic parametric interaction~\cite{supplement}.
The efficiency for other qubits are measured similarly and summarized in table S1 in Ref.~\cite{supplement}.

Ideally, LRU should not impact the computational subspace as long as there is no leakage population in the corresponding level, i.e., it is equivalent to an identity operation for states within the computational subspace~\cite{aliferis2005fault}.
In our implementations, the couplers are adiabatically tuned closer to the qubits to accelerate the parametric interaction~\cite{supplement}. Such operation may cause unwanted conditional phase accumulation~\cite{chu2021coupler}. The conditional phase can be compensated using an additional controlled phase gate, or be suppressed by adjusting the interaction frequency of the coupler. 
The single qubit phase accumulation resulting from the modulation can be readily corrected using virtual Z gates~\cite{mckay2017efficient}.
We perform simultaneous interleaved randomized benchmarking (RB)~\cite{magesan2012efficient} with $N=150$ cycles
of Clifford operations to quantify non-identity errors induced by the parallel data-qubit LRUs on the qubits, see Fig.~\ref{fig3}(c).
We note that to avoid crosstalk between qubits coupled to the same coupler, LRU for the ancilla qubit is not applied in parallel with the data qubits here, but tested in a separate RB experiment instead~\cite{supplement}.
Compared with the reference RB, we extract an error rate of $r_{\rm A}=0.82\%$ for qubit A, and $r_{\rm D1}=0.47\%$, $r_{\rm D2}=0.46\%$ for D1 and D2 respectively, with an average error rate of 0.58\% over three qubits, as shown in Fig.~\ref{fig3}(d). Error budget analysis, conducted by interleaving idle gates with the same duration, suggests that the majority of the error is attributed to decoherence~\cite{supplement}. Additional errors may stem from undesired transitions caused by parametric modulation or from undesired phase accumulation.
The results of the simultaneous data-qubit LRU demonstrated here prove the low crosstalk characteristics of our scheme. 

\begin{figure}
    \centering
    \includegraphics{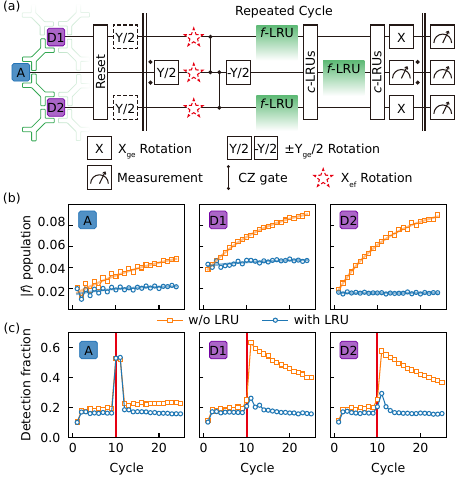}
    \caption{\label{fig4} Weight-two Z stabilizer measurement circuit incorporating the $f$-LRUs.
    (a) Quantum circuit using ancilla A to measure the Z-type parity of data qubits D1 and D2. An artificial leakage error is introduced by applying an $X_{ef}$ rotation to the corresponding qubit. A reset process is applied to all three qubits prior to the repeated circuit to initialize the system into the ground state. Echo ($X$) pulses are applied to the data qubits during the ancilla measurement.
    (b) The measured leakage population of each qubit versus the number of parity check cycles, with (blue) and without (orange) the $f$-LRUs, using a three-state readout. The data are averaged over 10 traces. The data qubits are initialized into a superposition state using $Y/2$ rotations.
    (c) The error-detection probability versus cycles, with an artificial leakage error injected to qubits A (left), D1 (middle), or D2 (right) at the 10-th cycle, with (blue) or without (orange) $f$-LRUs. The data are averaged over 10 traces.
    }
\end{figure}

Finally, we benchmark the performance of the $f$-LRUs within the surface code QEC context by incorporating it into repeated cycles of a weight-two Z-type stabilizer measurement~\cite{kelly2015state}, as depicted in Fig.~\ref{fig4}(a). 
Each cycle lasts 1.9 $\mu$s, including 1.024 $\mu$s for dispersive readout, 240 ns for CNOT gates, and the remaining time for LRUs and energy dissipation in the readout resonators.
We track the increase in leakage to the $|f\rangle$ state across the repeated cycles using a three-state readout for each qubit (see Fig.~\ref{fig4}(b)). The data qubits are initialized to a superposition state through $Y/2$ rotations. When $f$-LRUs are applied to the circuit, the accumulation of leakage is reduced by 0.026, 0.045, 0.074 after 25 cycles for qubit A, D1 and D2 respectively.
Note the nonzero intercept observed at 0-th cycle is mainly caused by readout errors~\cite{supplement}.
The odd-even oscillation observed in the ancilla qubit is due to the ancilla qubit's tendency to be in the $|e\rangle$ state during odd cycles, which leads to a higher measured rate of $|f\rangle$ because of readout error.
Our findings indicate that leakage to the $|f\rangle$ state of the data qubit is more pronounced, suggesting that two-qubit gates are the primary source of leakage errors, given the involvement of the $|f\rangle$ state in the two-qubit gate operations. Notably, significant leakage on the ancilla qubit is observed when the ancilla readout power is increased~\cite{sank2016measurement}, as detailed in the Supplementary Material~\cite{supplement}.

To illustrate the effectiveness of $f$-LRUs in mitigating time-correlated errors, we artificially introduce a leakage error by performing an $X_{ef}$ rotation between the $|e\rangle$ and $|f\rangle$ states on each qubit one-by-one during the 10-th cycle of a 25-cycle experiment.
Figure~\ref{fig4}(c) displays the fraction of error detection events, corresponding to the proportion of trials where a stabilizer measurement yields an unexpected outcome. The artificial leakage injected into the ancilla qubit and the data qubits result in distinct error syndromes: a pair of detection event fraction of 0.5 immediately after the injection into the ancilla, as the injection produces a random readout result; and a gradual decay of detection events over several cycles after injected into the data qubits, indicating temporally correlated errors. Even in the absence of artificial injection (cycles preceding the injection), a slow increase in error detection events is observed, suggesting intrinsic leakage-induced errors from the circuit itself.
With the $f$-LRUs inserted into the circuit, the error syndrome becomes more uniform, except for a few cycles immediately after the injection and the initial cycle due to initialization. Crucially, the previously observed slow decay of errors no longer exists.
These findings demonstrate that the $f$-LRUs are capable of significantly reducing leakage population on qubits, thereby effectively suppressing the time-correlated errors.

In summary, we have proposed and validated an efficient leakage reduction scheme for scalable QEC with superconducting qubits, leveraging the widely utilized tunable couplers. 
Capable of removing leakage on both the couplers and qubits, suppressing correlated errors in both space and time, our scheme demonstrates its potential as an indispensable building block for scalable QEC with superconducting qubits.
The limitation of our current work lies in the relatively slow dissipation rate of the couplers indirectly coupled to the broadband Purcell filter through the readout resonators.
Future improvements could be made by directly coupling the couplers to the Purcell filter, facilitating a significantly faster coupler dissipation rate without involving the readout resonators. This scheme should not compromise the performance of two-qubit gates, as the couplers remain outside the dissipative regime throughout the gate operations.

\begin{acknowledgments}
{
 This work was supported by the Science, Technology and Innovation Commission of Shenzhen Municipality (KQTD20210811090049034), the National Natural Science Foundation of China (12174178, 12204228, 12374474 and 123b2071), the Innovation Program for Quantum Science and Technology (2021ZD0301703), the Shenzhen-Hong Kong Cooperation Zone for Technology and Innovation (HZQB-KCZYB-2020050), and Guangdong Basic and Applied Basic Research Foundation (2024A1515011714, 2022A1515110615).}
\end{acknowledgments}

\bibliography{Reference}

\end{document}


\title{Supplemental Material for ``Coupler-Assisted Leakage Reduction for Scalable Quantum Error Correction with Superconducting Qubits"}
\maketitle

\setcounter{equation}{0}
\setcounter{figure}{0}
\setcounter{table}{0}
\setcounter{page}{1}

\tableofcontents
\newpage

\begin{figure}
    \centering
    \includegraphics{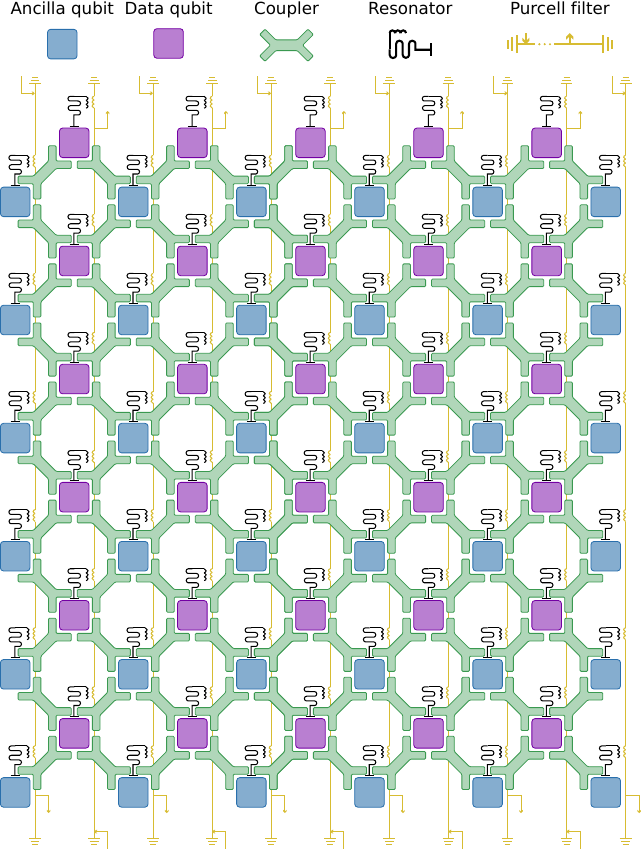}
    \caption{\label{fig_66qlayout} Schematic of the superconducting quantum processor with $6\times 11=66$ qubits (squares) and 110 tunable couplers (bone shape). The processor consists of two chips bonded together using flip-chip technology, where the qubits and couplers are positioned on the top chip, the readout resonators, Purcell filters and control wiring are patterned on the bottom carrier chip respectively.
    Each qubit has an individual readout resonator (black line) and each column of six resonators are coupled to the outside 50~$\Omega$ environment through a broadband Purcell filter (yellow line).
    }
\end{figure}

\section{Device and experimental setup}

\begin{figure}
    \centering
    \includegraphics{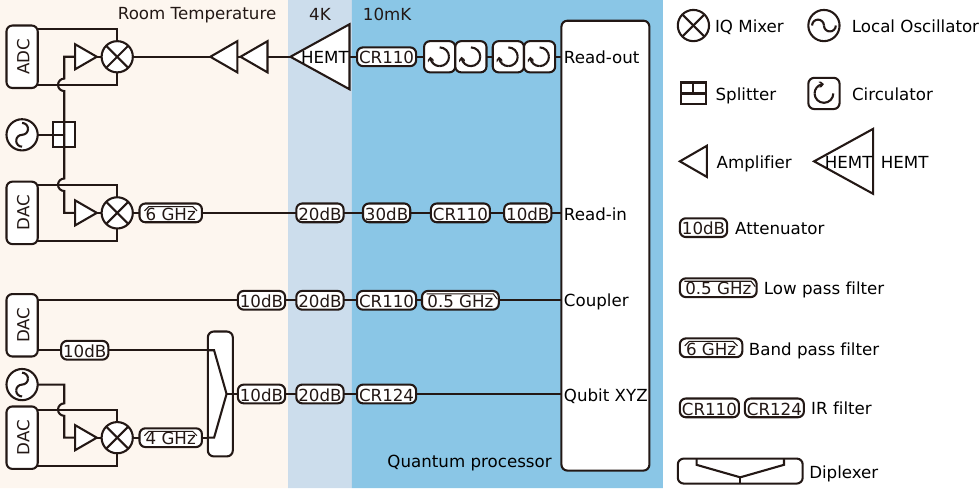}
    \caption{\label{sfig_setup} Schematic diagram of the experimental setup. See text for details.}
\end{figure}

The experiments discussed in this manuscript were conducted on a subset of three qubits and three tunable couplers in a superconducting quantum processor consisting of 66 qubits and 110 tunable couplers.
Figure~\ref{fig_66qlayout} illustrates the full diagram of the 66-qubit superconducting quantum processor. 
The processor consists of two chips which are bonded together using flip-chip technology~\cite{foxen2017qubit}: a top chip housing the qubits and couplers, and a bottom chip for the readout resonators, Purcell filters and control wiring.
Each qubit is connected to four nearest-neighboring qubits through tunable couplers. 
Each quarter-wavelength ($\lambda/4$) coplanar waveguide (CPW) readout resonator is capacitively coupled to its associated qubit at its open end and inductively coupled to a half-wavelength ($\lambda/2$) Purcell filter at its shorted end. Each Purcell filter is shared by six qubits and their corresponding readout resonators in the same column.
The Purcell filter is a bandpass filter between the resonator and the
50~$\Omega$ environment which impedes microwave propagation at the qubit frequency, while facilitating the microwave propagation at the readout frequency.
Different designs of Purcell filter have been devised, where the Purcell filter we use here is simply a half-wavelength CPW resonator with both ends shorted to ground, and the input and output lines intersect with the Purcell filter near the shorted ends.
The Purcell filter has a center frequency of about 6~GHz, with a weak coupling to the input port (coupling quality factor of approximately 2000) and a strong coupling to the output port (coupling quality factor of approximately 25).

The fabrication process of the quantum processor involves the following steps:
\begin{enumerate}
\item A 100 nm aluminum (tantalum) film is deposited onto a sapphire wafer for the bottom (top) chip respectively.
\item Large-scale structures, including the control and readout circuits on the bottom chip, as well as the capacitor pads for the qubits and couplers on the top chip, are realized through optical lithography and subsequent wet etching.
\item To mitigate signal crosstalk, SiO$_2$-supported bridges are created on the bottom chip to shield critical circuits.
\item The Al-${\rm AlO_x}$-Al Josephson junctions are patterned via electron beam lithography and fabricated using the double-angle electron beam evaporation.
\item Bandage technology~\cite{dunsworth2017characterization} is employed to establish a galvanic connection between the aluminum junctions and the tantalum film deposited in step 1.
\item 9-$\mu$m-tall SU-8 photoresist is positioned at the corners of the chips as a holder between the top and bottom chips, and then the top and bottom chips are bonded together using nLOF glue.
\end{enumerate}

\begin{table}
\begin{ruledtabular}
\begin{tabular}{lcccccc}
Component & ${\rm D1}$ & ${\rm C1}$ & ${\rm A}$ & C2  & D2 & C3\\
Maximum frequency, $\omega_{\textrm{max}}/2\pi (\mathrm{GHz})$ & 4.28 & $\sim 7.2$ & 4.27 & $\sim 7.2$ & 4.31 & $\sim 7.2$\\
Idle frequency, $\omega_{\textrm{idle}}/2\pi (\mathrm{GHz})$ & 4.172 & $\sim 5.5$ & 4.086 & $\sim 5.5$ & 4.239 & $\sim 5.5$\\
Anharmonicity, $\alpha/2\pi$ ($\mathrm{MHz}$) & $-211$ & $\sim-135$ & $-208$ & $\sim -135$ & $-205$ & $\sim -135$\\
Readout frequency, $\omega_r/2\pi$ ($\mathrm{GHz}$) & 5.886 &  & 6.039 &   & 5.858 &  \\
Resonator linewidth, $\kappa_{\rm r}/2\pi$ ($\mathrm{MHz}$) & 5.4 &   & 3.0 &   & 5.9 & \\
Dispersive shift, $2\chi/2\pi$ ($\mathrm{MHz}$) & $-2.9$ &   & $-1.9$ &   & $-4.0$ &  \\
Single qubit gate error, $\xi_{1Q}$ ($\%$) & 0.07 &   & 0.08 &   & 0.07 &  \\
Two qubit gate error, $\xi_{2Q}$ ($\%$) &   & 0.78 &   & 0.99 &   &   \\
$T_1$ at idle frequency ($\mu$s) & 62 & $\sim 14$ & 71 & $\sim 10$ & 67 & $\sim 8$ \\
$T_2^*$ at idle frequency ($\mu$s) & 2.69 & - & 4.40 & - & 3.07 & - \\
$T_{2,\rm echo}$ at idle frequency ($\mu s$) & 22.69 & - & 8.02 & - & 30.94 & -\\
Modulation frequency of $e$-Reset ($\mathrm{MHz}$) &   & 618 &   & 650 &   & 559\\
Modulation frequency of $f$-LRU ($\mathrm{MHz}$) &   & 777 &   & 706 &   & 722\\
Modulation frequency of $h$-LRU ($\mathrm{MHz}$) &   & - &   & 505 &   & - \\
Duration of $e$-Reset (ns) & 76 & & 75 & & 77 \\
Duration of $f$-LRU (ns) & 90 &   & 98 &   & 96 &  \\
Coupler-Resonator iSWAP (ns) &   & 9 &   & 9 &   & 9\\
Efficiency of $f$-LRU (\%) & 95.4 &   & 98.1 &   & 98.1 &  \\
Efficiency of $h$-LRU (\%) & - &   & 95.1 &   & - &   \\
\end{tabular}
\end{ruledtabular}
\caption{\label{tabel_qubits} Parameters, coherence properties, and gate errors of the qubits, as well as the parameters for the leakage reduction operations described in the main text.
}
\end{table}

The processor is wire-bonded to a printed circuit board enclosed by aluminum.
The packaged processor is mounted to the mixing chamber stage (with a base temperature down to $\sim 10$~mK) of a dilution refrigerator and connected to room-temperature electronics through attenuators, filters and amplifiers, as summarized in Fig.~\ref{sfig_setup}.

Microwave signals for single-qubit XY control and dispersive readout are up-converted from carriers generated by a local oscillator using IQ mixing. 
Intermediate frequency signals for up-convertion and for qubit/coupler flux bias are generated by arbitrary waveform generators (AWGs) with a sampling rate of 2.0 Gsa/s (HDAWG from Zurich Instrument). Diplexers are employed to combine the XY signal and the qubit Z signal at room temperature.
Inside the refrigerator, attenuators and infra-red filters are positioned along the control lines at different stages for thermalization and noise attenuation. 
The readout signals from the quantum processor are amplified by high electron mobility transistor (HEMT) amplifiers at the 4~K stage, followed by further amplification at room temperature. Prior to amplification, circulators and infrared filters are installed at the mixing chamber stage to block noise originating from higher temperature stages. The output signals are subsequently down-converted to intermediate frequency signals, digitized and demodulated using the FPGA-based acquisition unit of the UHFQA from Zurich Instruments.

Table~\ref{tabel_qubits} provides an overview of the parameters for the qubits and couplers utilized in this experiment.

\section{Stray interaction}

\begin{figure}
    \centering
    \includegraphics{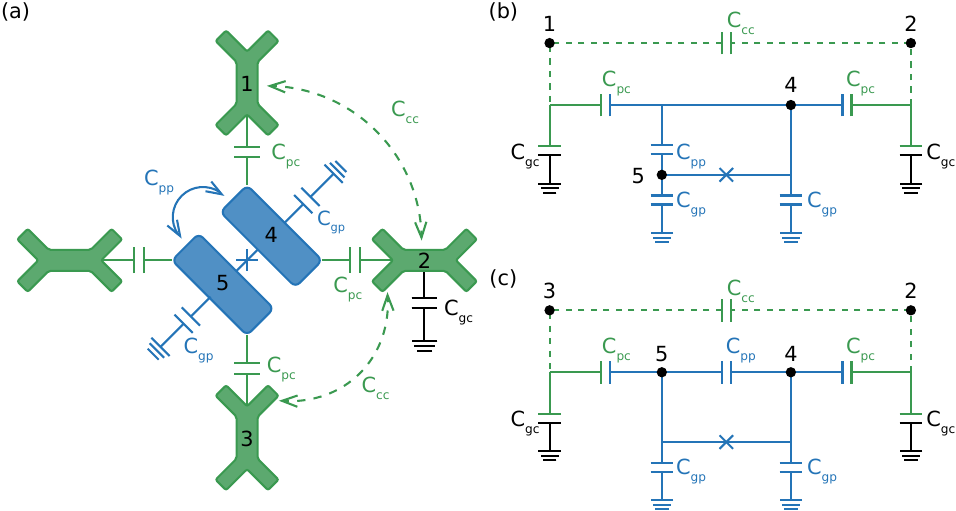}
    \caption{\label{fig:coupler_stray} 
    Schematic diagram of stray interaction between neighboring couplers. (a) Sketch of a floating transmon qubit (blue) surrounded by four tunable couplers (green). (b) Circuit diagram illustrating the capacitance network in the case of asymmetric coupling, where two couplers are coupled to the same pad of the qubit shown in (a). (c) Circuit diagram illustrating the capacitance network in the case of symmetric coupling, where two couplers are coupled to different pads of the qubit.
    }
\end{figure}

\subsection{Stray interaction between neighboring couplers}

The qubits used in our sample are floating transmons, which is now a prevalent choice in two-dimensional superconducting chips~\cite{arute2019quantum,wu2021strong}. The tunable couplers are transmon qubits as well, except that they have higher maximum frequencies and weaker anharmonicity compared to the qubits.
Consider a floating qubit surrounded by four couplers, as depicted in Fig.~\ref{fig:coupler_stray}. We use the letter 'p' to refer to the pads of the qubit. Specifically, $C_{\rm pc}$ represents the coupling capacitance between a qubit pad and a coupler pad. $C_{\rm pp}$ is the capacitance between two pads of the floating qubit, and $C_{\rm gp}$ denotes the capacitance between a qubit pad and the ground. Additionally, $C_{\rm gc}$ and $C_{\rm cc}$ represent the coupler-to-ground and coupler-to-coupler capacitances, respectively. 
The stray coupling between neighboring couplers can be categorized into two cases: asymmetric coupling, as depicted in Fig.~\ref{fig:coupler_stray}(b), where two couplers are coupled to the same pad of the qubit, and symmetric coupling, as depicted in Fig.~\ref{fig:coupler_stray}(c), where two couplers are coupled to different pads of the qubit.

Following Ref.~\cite{yan2018tunable,sete2021floating}, we derive the capacitive coupling between the qubit and the couplers. 
The capacitive coupling between neighboring couplers differs in the symmetric and asymmetric case, as
\begin{equation}
\begin{aligned}
g^{\rm a}_{\rm cc} &\approx  \frac{1}{4} [\frac{C_{\rm pc}^2 (C_{\rm gp} + C_{\rm pp}) } {C_{\rm gp}(C_{\rm pp} +C_{\rm gp}/2 ) C_{\rm gc} } + \frac{2C_{\rm cc}}{C_{\rm qc}} ] \cdot \omega_{\rm c}, \\
g^{\rm s}_{\rm cc} &\approx \frac{1}{4} [\frac{C_{\rm pc}^2  C_{\rm pp} } {C_{\rm gp}(C_{\rm pp} +C_{\rm gp}/2  )C_{\rm gc} } + \frac{2C_{\rm cc}}{C_{\rm qc}} ] \cdot \omega_{\rm c}.
\end{aligned} \label{eq:stray_asymm}
\end{equation}
In the above expressions, some small terms are neglected assuming that $C_{\rm gc},C_{\rm gp} \gg C_{\rm pp},C_{\rm pc},C_{\rm cc}$.
The qubit-coupler coupling is given by
\begin{equation}
\begin{aligned}
g_{\rm q,c} &\approx  \frac{1}{4} \frac{C_{\rm pc}}{ \sqrt{(C_{\rm pp} + C_{\rm gp}/2)C_{\rm gc}}  } \cdot \sqrt{\omega_{\rm q} \omega_{\rm c}}.
\end{aligned} \label{eq:stray_asymm2}
\end{equation}
When two couplers are coupled to the different (the same) pads of a qubit, the qubit-coupler coupling has the same (the opposite) signs~\cite{sete2021floating}.
Therefore, the direct capacitive stray coupling and the indirect coupling mediated by the qubit can interfere constructively or destructively.
In the asymmetric case as shown in Fig.~\ref{fig:coupler_stray}(b), they interfere constructively and yield a total stray coupling of
\begin{equation}
\widetilde{g}_{\rm cc}^a = g_{\rm cc}^a  + \frac{g_{\rm q,c}^2}{\omega_{\rm c} - \omega_{\rm q}},
\end{equation}
whereas in the symmetric case as exemplified in Fig.~\ref{fig:coupler_stray}(c), the total stray coupling is quantified by
\begin{equation}
\widetilde{g}_{\rm cc}^s = g_{\rm cc}^s  - \frac{g_{\rm q,c}^2}{\omega_{\rm c} - \omega_{\rm q}}.
\end{equation}
Electromagnetic simulations of our layout pattern suggests that $C_{\rm gc}=150$~pF, $C_{\rm gp} = 140$~fF, $C_{\rm pp}=20$~pF, $C_{\rm pc}=11$~pF, $C_{\rm cc} = 0.04$~fF respectively.
At the idling point, $\omega_c/2\pi\approx 5.5$~GHz, $\omega_q/2\pi\approx 4.2$~GHz, 
we estimate that $\widetilde{g}_{\rm cc}^a/2\pi\approx 11$~MHz and $\widetilde{g}_{\rm cc}^s/2\pi\approx -23$~MHz, with an average (absolute) stray coupling of $\sim 17$~MHz over the whole device.
For the specific case of the three couplers C1, C2, C3 used in this experiment, the stray coupling between C1 and C2, C2 and C3 are both asymmetric.

Theoretically, it is possible to tune the stray coupling between neighboring couplers to zero in the symmetric case. However, achieving such cancellation in a two-dimensional quantum processor is challenging due to two reasons: (1) In the symmetric case, the direct coupling between neighboring couplers is typically weak ($\sim 1$~MHz). As shown in Eq.\ref{eq:stray_asymm}, the direct coupling in the symmetric case is much smaller, since $C_{\rm pp} \ll C_{\rm gp} + C_{\rm pp}$. The direct capacitance $C_{\rm cc}$ is negligible unless intentionally enhanced in the layout. 
Therefore, the weak direct coupling between couplers is insufficient to cancel out the indirect coupling mediated by the qubit mode. 
Increasing the detuning $\omega_{\rm c} - \omega_{\rm q}$ between the coupler and qubit, though potentially effective, goes against the original design intention of deactivating qubit-qubit coupling. (2) In a two-dimensional superconducting quantum processor, it is difficult, if not impossible, to ensure that all neighboring couplers are coupled either symmetrically or asymmetrically. Suppressing stray coupling between couplers is challenging due to the large number of couplers (twice the number of qubits) and the higher priority of suppressing stray coupling between qubits.

Fortunately, prior research has revealed that stray coupling between couplers does not significantly increase two-qubit gate error~\cite{chu2021coupler}. However, the leakage propagation among these couplers, triggered by such stray coupling, poses a considerable hurdle for quantum error correction, as we emphasized in the main text.

\subsection{Stray coupler-resonator interaction}

\begin{figure}
    \centering
    \includegraphics{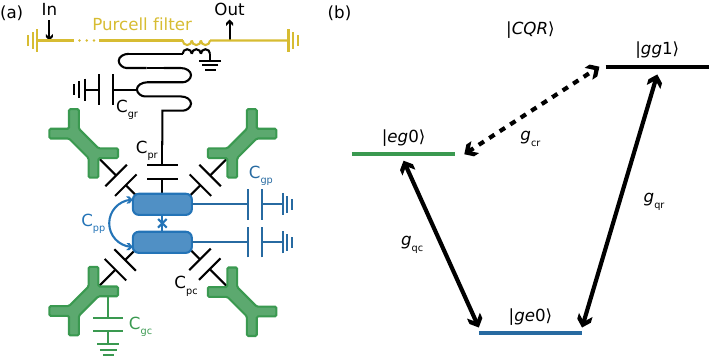}
    \caption{\label{sfig_coupling} 
    Schematic diagram of stray interaction between a coupler and a resonator.
    (a) The capacitance network surrounding a floating qubit.
    The readout resonator is a quarter-wavelength CPW resonator, with the open end capacitively coupled to the qubit and the other shorted end inductively coupled to the Purcell filter. The effective capacitance of the resonator is $C_{\rm r}$.
    (b) The level diagram of the $\rm |coupler,qubit,resonator\rangle$ ($\rm CQR$) system in the one-excitation manifold. Solid lines with arrows represent nearest-neighbour coupling, while the dashed line with arrows represents the indirect capacitive coupling through the intermediate capacitance network.
    }
\end{figure}


In the quantum processor, each readout resonator is a quarter-wavelength meandered coplanar waveguide (CPW) resonator, with the open end capacitively coupled to a qubit and the other shorted end inductively coupled to the Purcell filter, as shown in Fig.~\ref{sfig_coupling}(a).
Similar to the stray coupling between neighboring couplers, the stray coupler-resonator interaction also has two types: the symmetric case where the coupler and the resonator couple to different qubit pads and the asymmetric case where the coupler and the resonator couple to the same qubit pad. 
As depicted in Fig.~\ref{sfig_coupling}(b), the net coupling between the resonator mode and the coupler mode, denoted as $\widetilde{g}_{\rm cr}$, is comprised of both direct capacitive coupling and indirect coupling mediated by the qubit mode, as
\begin{equation}
\begin{aligned}
        \widetilde{g}_{\rm cr} &= g_{\rm cr} + \frac{1}{2} g_{\rm qc}g_{\rm qr} (\frac{1}{\Delta_{\rm cq}} + \frac{1}{\Delta_{\rm rq}} ).
\end{aligned}
\end{equation}
Both the coupler-qubit detuning, $\Delta_{\rm cq} = \omega_{\rm c} - \omega_{\rm q}$, and the resonator-qubit detuning, $\Delta_{\rm rq} = \omega_{\rm r} - \omega_{\rm q}$, are positive.
The direct and indirect coupling add construtively (destructively) in the asymmetric (symmetric) case. To accelerate the $c$-LRU process, we always choose the asymmetric coupler-resonator interaction in our experiments.
The numerical simulation of our device layout suggests that $\widetilde{g}_{\rm cr}/2\pi\sim 25$~MHz in the asymmetric case, assuming $\omega_{\rm c}/2\pi,\omega_{\rm r}/2\pi = 6$~GHz and $\omega_{\rm q}/2\pi = 4.2$~GHz.
Experimental calibration shows that the actual $\widetilde{g}_{\rm cr}/2\pi\sim 30$~MHz, close to the numerical simulation result, see \rfig{sfig_calibration}(f) for details.
The implementation of the leakage reduction scheme relies on the nonzero stray coupler-resonator coupling $\widetilde{g}_{cr}$, allowing for rapid dissipation of the coupler population through the resonator strongly coupled to the Purcell filter. 
As mentioned in the conclusion part of the main text, future improvements could be made if directly couple the coupler to the Purcell filter.
Importantly, this scheme should not compromise the performance of two-qubit gates, as the couplers remain outside the dissipative regime throughout the gate operations.

\section{Multi-level transmon readout and Gaussian mixture model}{\label{GMM}}

\begin{figure}
    \centering
    \includegraphics[width=1.0\linewidth]{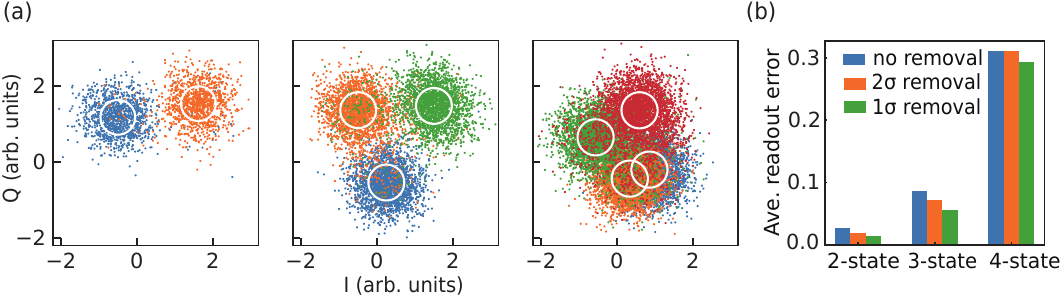}
    \caption{\label{sfig_readout_error} 
    Multi-level readout of the ancilla qubit A.
    (a) Single-shot 2-state (left), 3-state (middle), 4-state (right) readout data for the lowest energy states of A. The data plots show the initial 2000 shots out of a total of $2^{12}$ shots for each state. White circles indicate the standard deviation of the Gaussian distribution for the respective states. The centers and variances of each state are determined using the Gaussian mixture model.
    (b) The average assignment errors for 2-state, 3-state, and 4-state readouts, without outlier removal, are 2.6\%, 8.5\%, and 31.1\%, respectively. Upon excluding data points outside of 2 (1) standard deviations, the corresponding assignment errors are reduced to 1.8\% (1.3\%), 7.1\% (5.5\%), and 31.0\% (29.3\%), respectively.
    }
\end{figure}

In a typical circuit quantum electrodynamics setup~\cite{blais2021circuit}, the transmon qubit is read out by probing the state-dependent dispersive shift of its readout resonator. The probe signal is amplified, digitized, and then demodulated, yielding a data point in the in-phase/quadrature (IQ) phase plane that depends on the phase shift imparted by the qubit.
Taking $n$ energy levels of the transmon into account, and repeating the single-shot readout of the transmon qubit many times, we can obtain a scatter plot in the IQ plane, see Fig.~\ref{sfig_readout_error} for example, where each qubit state corresponds to a 2D Gaussian distribution in the IQ plane. We employ an $n$-component Gaussian mixture model to estimate the centers and variances for all the $n$ states, denoted as $c_i$ and $\sigma_i$ for the state $|i\rangle$, respectively.
The probability that a single-shot readout data point $v$ belongs to the Gaussian distribution of state $|i\rangle$ is calculated using the formula
\begin{equation}
    p_{i,v} = \frac{1}{2\pi \sigma_i^2} e^{-\frac{(v - c_i)^2}{2\sigma_i^2}}.
\end{equation}
We assign a single-shot data $v$ to state $|s(v)\rangle$ using
\begin{equation}
    s(v) =
    \begin{cases}
      -1, & \text{if } \max_{i \in \mathcal{S}_n} p_{i,v} < p_{\text{th}}, \\
      {\rm argmax}_{i \in \mathcal{S}_n } p_{i,v}, & \text{otherwise.}
    \end{cases}
\end{equation}
Here, $\mathcal{S}_n = \{g, e, f, \ldots\}$ denotes the set comprising the relevant $n$ quantum states. The function $s$ is designed to map raw data onto state indices. We define any single-shot data point that falls outside $k\sigma$ standard deviations from the Gaussian distributions as an outlier, which is indicated by $-1$. This classification is based on the threshold $p_{\rm th} = \frac{1}{2\pi \sigma_g^2}e^{-\frac{k^2}{2}}$, where the variance of the ground state ($\sigma_g^2$) is utilized as the reference.
For a readout data trace $V$, the probability of the qubit being in state $|i\rangle$ is determined by:
\begin{equation}
    P_{i} = \frac{{\rm count}(s(v) = i, v \in V)}{{\rm count}(s(v) \geq 0, v \in V)}.
\end{equation}

Figure~\ref{sfig_readout_error}(a) displays the single-shot readout data for 2-state (left), 3-state (middle), and 4-state (right) readout. State discrimination is performed using the model outlined above. The average assignment error $\epsilon_n$ for an $n$-state readout is defined by the following equation:
\begin{equation}
    \epsilon_n = 1 - \frac{1}{n}\sum_{i \in \mathcal{S}_n} P_{ii},
\end{equation}
where $P_{ii}$ represents the measured probability of observing state $|i\rangle$ when the qubit is prepared in the same state $|i\rangle$.
Figure~\ref{sfig_readout_error}(b) illustrates the assignment errors both with and without the removal of outliers. 
The exclusion of outliers minimizes assignment errors by reducing separation errors.
Elevating the outlier threshold $p_{\rm th}$ (which corresponds to reducing $k$) enhances readout fidelity, albeit at the cost of excluding more data points.
Given the relatively low readout signal-to-noise ratio in our experiments because of the absence of parametric amplifiers, we have opted to exclude data points lying outside one standard deviation in the data analysis featured in Fig.~3(b) and Fig.~4(b-c) of the main text.

\section{Coupler-assisted reset and leakage reduction}

\subsection{Parametric modulation of the coupler frequency}

\begin{figure}
    \centering
    \includegraphics[width=0.5\linewidth]{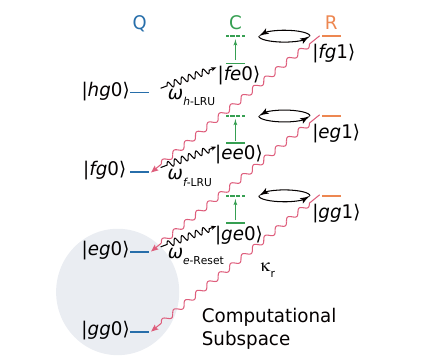}
    \caption{\label{sfig_schematic}
    Energy level spectrum of the QCR system. The levels are labeled as $|\rm QCR\rangle$, where the number indicates the number of photons in the resonator.
    The computational subspace (gray area) consists of the states $|gg0\rangle$ and $|eg0\rangle$. All states outside the computational subspace are considered leakage states. The qubit reset and leakage reduction scheme involves transferring the population from higher energy qubit levels to the coupler by parametrically modulating the coupler at specific frequencies ($\omega_{e-\rm Reset}$, $\omega_{f-\rm LRU}$, $\omega_{h-\rm LRU}$), represented by the black 
   wave lines. The coupler is subsequently reset to its ground state by tuning it into resonance with the readout resonator, a dissipative element where the excitations decay quickly at rate $\kappa_{\rm r}$, as indicated by the red wave lines.
    }
\end{figure}

In our leakage reduction scheme, the readout resonator serves as a sink for energy dissipation, whereas the coupler facilitates the transfer of excitations from the qubit to the resonator, meanwhile removing leakage population on itself. The energy level spectrum of such a qubit-coupler-resonator (QCR) system is depicted in Fig.~\ref{sfig_schematic}. 
State transfer between the qubit and the coupler is realized by parametrically modulating the tunable coupler at a specific frequency $\omega_p$~\cite{chu2023scalable,caldwell2018parametrically,reagor2018demonstration}.
The modulation frequencies for qubit $|e\rangle$ state reset (referred to as $e$-Reset), $f$-LRU, $h$-LRU processes are given by:
\begin{equation}\label{eq_param_freq}
	\left\{
	\begin{aligned}
		m \cdot \omega_{e\rm-Reset}  &= |\omega_{ge} - \bar{\omega}_{\rm c}|, \\
		m \cdot \omega_{f\rm-LRU}  &= |\omega_{ef} - \bar{\omega}_{\rm c}| \approx |\omega_{ge} + \alpha - \bar{\omega}_{\rm c}|, \\
		m \cdot \omega_{h\rm-LRU}  &= |\omega_{fh} - \bar{\omega}_{\rm c}| \approx |\omega_{ge} + 2\alpha - \bar{\omega}_{\rm c}|,
	\end{aligned}
	\right.
\end{equation}
where $m$ denotes the $m$-th harmonic interaction in parametric control, $\omega_{ij}$ is the transition frequency between the qubit states $|i\rangle$ and $|j\rangle$, $\alpha$ is the anharmonicity of the transmon qubit, and $\bar{\omega}_{\rm c}$ is the time-averaged frequency of the coupler during the parametric modulation. The effective coupling strength $g$ between the target levels is determined by the modulation amplitude $A_{\rm p}$~\cite{caldwell2018parametrically}, 
\begin{equation}\label{eq_g}
    g = \sqrt{n_{\rm ex}} g_{\rm qc} {\rm J}_{m}(\frac{A_{\rm p}}{m \omega_{\rm p}}),
\end{equation}
where $g_{\rm qc}$ is the intrinsic coupling strength between the qubit and the coupler, $n_{\rm ex}$ is the total excitations of the corresponding level, ${\rm J}_{m}(\cdot)$ the $m$-th Bessel function of the first kind. Once an excited state is selectively transferred to the coupler, the coupler is reset by tuning it into resonance with the dissipative resonator. 
The leakage reduction or reset process is considered complete when the excitations on both the coupler and resonator have  undergone sufficient decay.



\begin{figure}
    \centering
    \includegraphics{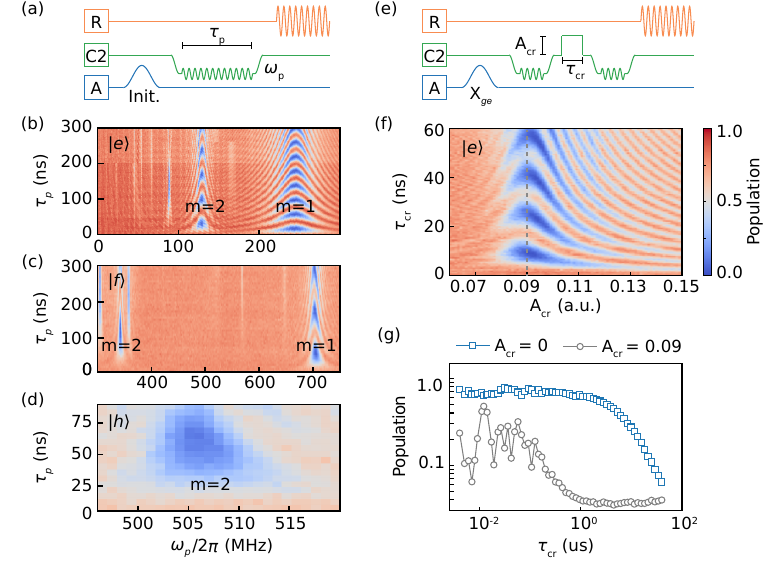}
    \caption{\label{sfig_calibration}
    Calibration of the qubit-coupler (QC) swap and coupler-resonator (CR) interaction for qubit A and coupler C2.
    (a) Pulse sequence for calibrating the QC swap, including a microwave pulse for qubit state initialization (blue), a parametric pulse with adiabatic edges on the coupler (green), and a measurement pulse (orange).  The qubit is initialized to $|e\rangle$, $|f\rangle$, and $|h\rangle$ for the $e$-Reset, $f$-LRU, and $h$-LRU operations, respectively.
    (b) Parametrically activated swap between $|eg0\rangle$ and $|ge0\rangle$ for $e$-Reset, showing the change in $|e\rangle$ state population with parametric pulse frequency and duration. The two chevron patterns represent the first ($m=1$) and second ($m=2$) order harmonic interactions as described in Eq.~\ref{eq_param_freq}.
    (c) Parametrically activated swap between $|fg0\rangle$ and $|ee0\rangle$ for $f$-LRU. 
    (d) Parametrically activated swap between $|hg0\rangle$ and $|fe0\rangle$ for $h$-LRU with the $|f\rangle$ state population been shown. 
    (e) Pulse sequence for calibrating the CR swap. The coupler is initialized using an $X_{ge}$ gate on an adjacent qubit, followed by a calibrated QC swap. A square pulse with amplitude $A_{cr}$ and duration $\tau_{cr}$ is applied to tune the coupler into resonance with the resonator. A subsequent QC swap transfers the population from the coupler back to the qubit for measurement.
    (f) Measured population of coupler $|e\rangle$ versus  $A_{cr}$ and $\tau_{cr}$. The dashed gray line indicates the resonant swap between the coupler and the resonator.
    (g) Accelerated dissipation of the coupler under resonance conditions, in comparison to the idle state ($A_{cr} = 0$, blue line). 
    Subplots (b), (c), (d), and (f) share the same colorbar located on the right side of (f).
    }
\end{figure}

The calibration process for the reset/LRU scheme includes two stages: the state swap between the qubit and the coupler, as shown in Fig.~\ref{sfig_calibration}(a-d), and coupler dissipation by tuning it into resonance with the resonator, as shown in Fig.~\ref{sfig_calibration}(e-g).
The experimental sequence for the qubit-coupler (QC) swap is shown in Fig.~\ref{sfig_calibration}(a), including a microwave pulse for qubit state initialization (blue), a parametric pulse with adiabatic edges on the coupler (green), and a measurement pulse (orange). The qubit is initialized to $|e\rangle$, $|f\rangle$, and $|h\rangle$ for the $e$-Reset, $f$-LRU, and $h$-LRU operations, respectively.
The result of the parametrically activated swap for $e$-Reset and $f$-LRU is shown in Fig.~\ref{sfig_calibration}(b) and (c), with the two chevron patterns representing the first ($m=1$) and second ($m=2$) order harmonic interactions described in Eq.~\ref{eq_param_freq}. 
For the $h$-LRU, only the second harmonic interaction is accessed, as shown in Fig.~\ref{sfig_calibration}(d), since the first harmonic interaction is beyond the bandwidth of the control electronics. The subdued visibility of data in the $h$-LRU is due to the low signal-to-noise ratio associated with the four-state readout, as shown in Fig.~\ref{sfig_readout_error}.
The calibration parameters for $e$-Reset, $f$-LRU, and $h$-LRU correspond to the points of minimum occupancy in the $|e\rangle$, $|f\rangle$ and $|h\rangle$ state in Fig.~\ref{sfig_calibration}(b)-(d) respectively.
In this work, the duration of each parametric pulse is maintained at approximately 60 ns.

A notable benefit of our scheme is the relaxation on the bandwidth of the control electronics, whose price dramatically surges as the bandwidth increases.
The coupler frequency here in idle status (deactivating coupling between adjacent qubits) is largely detuned ($>1$~GHz) from the qubit frequency.
Direct parametric modulation would require DAC channels with a large bandwidth.
Alternatively, employing a higher-order interaction ($m>1$) results in a considerably weaker interaction, as described by Eq.\ref{eq_g}.
Leveraging the strong tunability of the coupler, our method allows for adiabatic tuning of the coupler close to the qubits and the subsequent application of parametric pulses, thus relaxing the bandwidth requirement of the control electronics.

The coupler-resonator swap is realized by tuning the coupler frequency into resonance with the readout resonator using a square pulse, as depicted in Fig.~\ref{sfig_calibration}(e).
The coupler is initialized with an $X_{ge}$ gate on an adjacent qubit, followed by a calibrated QC swap, and indirectly measured by swapping the population back to the qubit.
The measured occupation rate of the coupler is shown in Figure~\ref{sfig_calibration}(f). A chevron pattern is observed on a time scale of tens of nanoseconds, where the resonant condition is indicated by the dashed line in Fig.~\ref{sfig_calibration}(f).
We prolong the pulse duration under resonant condition and observe a damped oscillation in Fig.~\ref{sfig_calibration}(g), since the coupling strength $g_{cr}$ between the coupler and the resonator is much larger than the decay rate $\kappa_{\rm r}$ of the resonator in our device.  
We can remove the population on the coupler using two different ways: swap the excitation from the coupler to the resonator ($\sim 9$~ns), where it decays at the intrinsic resonator decay rate $\kappa_{\rm r}$, or hold the square pulse for sufficient time until both qubit and resonator dissipate with the joint decay rate $\kappa_{\rm r}/2$~\cite{huber2024parametric}. 
While the latter is simpler in control, it comes at the cost of a slower dissipation rate. Considering the relative short dephasing time of the qubits, we choose to use the former method to implement the $c$-LRUs in Fig.~3 and Fig.~4 in the main text. The idle time for resonator decay in Fig.~3(c) is 50 ns. There is sufficient time for resonator decay in the stabilizer measurement circuit in Fig.~4, due to the staggered arrangement of the data LRUs and ancilla LRU.


\subsection{Coupler-assisted qubit reset} 

\begin{figure}[H]
    \centering
    \includegraphics{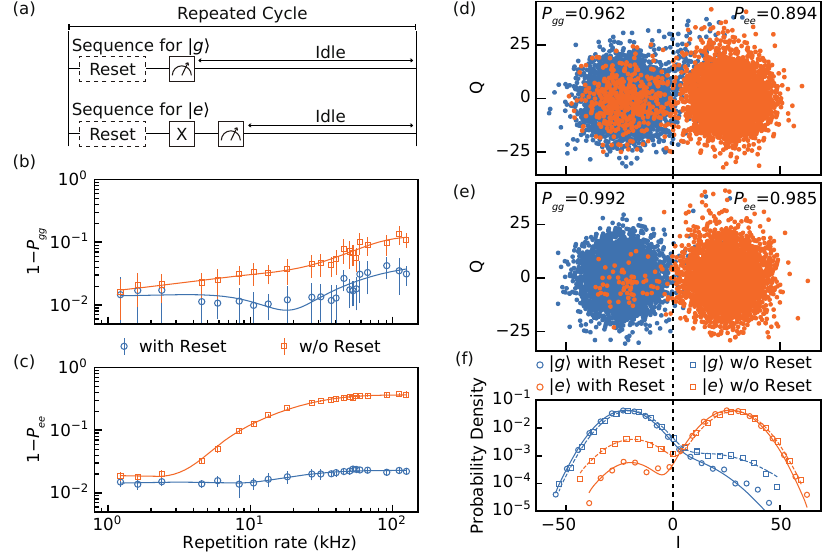}
    \caption{\label{sfig_readout_trigger} Qubit readout fidelity improvement by the Reset process. 
    (a) Sequence schematic for measuring readout errors at various repetition rates, which determines the interval between consecutive experiments, with or without applying the Reset process.
    (b) and (c) depict the readout errors for qubits initialized to $|g\rangle$ and $|e\rangle$ across various repetition rates, with (blue, circle) and without(orange, cross) applying the Reset process. 
    Each data point is averaged from 4000 raw data. 
    (d) and (e) display the single-shot readout data in the IQ panel without and with Reset process respectively.
    (f) Histogram of the IQ data in (d) and (e).
    }
\end{figure}

As mentioned in the main text, process similar to the leakage reduction operations for higher qubit states can be applied to reset the qubit $|e\rangle$ state as well, increasing the rate at which experiments are carried out by initializing the qubit to its ground state before the start of the next experiment.
Here, we present the details of the reset process and demonstrate its efficiency in improving the qubit readout fidelity when increasing the experiment repetition rate.
To avoid confusions, we clarify the distinction between "$e$-Reset" and "Reset" here: the former refers to the swap operation between the qubit and the coupler, while the latter encompasses the $e$-Reset process and the simultaneous $c$-LRUs process before and after the $e$-Reset process. The simultaneous $c$-LRUs preceding the $e$-Reset are applied to remove possible thermal excitations on the couplers.
The pulse sequence is shown in Fig.~\ref{sfig_readout_trigger}(a), with the experiment repetition rate dictating the interval between consecutive experiments.
Fig.~\ref{sfig_readout_trigger}(b) and (c) present the readout errors for $|g\rangle$  and $|e\rangle$ across various repetition rates respectively, both with (blue) and without (orange) the Reset process.
Without the Reset, the readout error of $|e\rangle$ state $1-P_{ee}$ becomes significant ($>10\%$) at a high repetition rate ($>10$~kHz). This is due to the necessity for a sufficient delay, typically 10 to 20 times the qubit's $T_1$ time, to allow the excited state to dissipate.
By employing the Reset to bring the qubit to the ground state, experiments can be conducted at elevated repetition rates (up to 100~kHz) while maintaining minimal readout error (below 2\%).
For $|g\rangle$ state, an increase in readout error $1-P_{gg}$ is also observed at higher repetition rates, which we attribute to readout-induced excitation~\cite{sank2016measurement}.
For both $|g\rangle$ and $|e\rangle$ states, the Reset process enhances readout fidelity even at low repetition rates, as it also mitigates readout errors induced by thermal excitations.

The improvement in readout fidelity due to the Reset is evident in Fig.~\ref{sfig_readout_trigger}(d)-(f). With a typical repetition rate of 10~kHz, corresponding to a 100~$\mu$s delay between experiments, the two-state readout fidelity reaches 98.9\% when Reset is applied, compared to 92.8\% without it.
This outcome is particularly noteworthy given that no parametric amplifier is utilized in this research. Note that this readout fidelity is obtained from readout data processed with the Gaussian mixture model (GMM) described in Section.~\ref{GMM}, without any outlier removal.

\subsection{$h$-LRU efficiency}
The definition for the efficiency of $h$-LRU is given by:
\begin{equation}
    \eta_h = 1 - \frac{P_{hh,\mathrm{LRU}}-P_{hf}}{P_{hh}}.
\end{equation}
Here, $P_{hf}$ represents the measured population of $|h\rangle$ with the qubit initially prepared in $|f\rangle$. This excitation mainly originates from readout-induced excitation. On the other hand, $P_{hh,\rm LRU}$ represents the measured population of $|h\rangle$ when the qubit is initially prepared in $|h\rangle$ and the $h$-LRU is applied.

\section{Error budget} %

\subsection{Interleaved randomized benchmarking}
    Randomized Benchmarking (RB)~\cite{magesan2012efficient} is a widely adopted method to characterize gate errors. It consists of a reference RB sequence and an interleaved RB sequence, performend in the following way:
    
    (i) The reference sequence or the standard randomized benchmarking sequence~\cite{knill2008randomized}: $m$ random Clifford gates are performed and then a unique recovery Clifford gate is appended to invert the operation of the Cliffords ahead and yield an identity if there is no error.
    The population of the grounded state, denoted as the sequence fidelity, decays exponentially due to Clifford errors:   
    \begin{equation}
        F_{\rm ref}(m) = Ap_{\rm ref}^m + B.
    \end{equation}
    Here $F_{\rm ref}(m)$ is the reference sequence fidelity, $m$ is the number of Clifford gates and $p_{\rm ref}$ is the depolarizing parameter.
    The average error rate per Clifford of the reference is 
    $r_{\rm ref} = (1 - p_{\rm ref})(d-1)/d$, with $d = 2^{N_{\rm q}}$ is the dimension of the system. The number of qubits $N_{\rm q}$ equals to 1 in our cases.

    (ii) The interleaved sequence differs from the reference by inserting the target operation, the LRUs in our experiments, behind every Clifford gates. The sequence fidelity decays exponentially due to Clifford errors as well as errors from the interleaved operation:  
    \begin{equation}
        F_{\rm int}(m) = Ap_{\rm int}^m + B.
    \end{equation}
    Here $F_{\rm int}(m)$ is the sequence fidelity of the interleaved sequence, and $p_{\rm int}$ is the depolarizing parameter.
    The average error rate per Clifford of the interleaved sequence $r_{\rm int} = (1 - p_{\rm int})(d-1)/d =r_{\rm ref} + r_{\rm LRU}$ includes both the average Clifford error the reference and the extra error induced by the LRUs. Therefore, we infer the error on the specific qubit caused by the LRUs using $ r_{\rm LRU} = r_{\rm int} - r_{\rm ref} $.

\subsection{Error budget for LRUs}

\begin{figure}[H]
    \centering
    \includegraphics[width=1.0\linewidth]{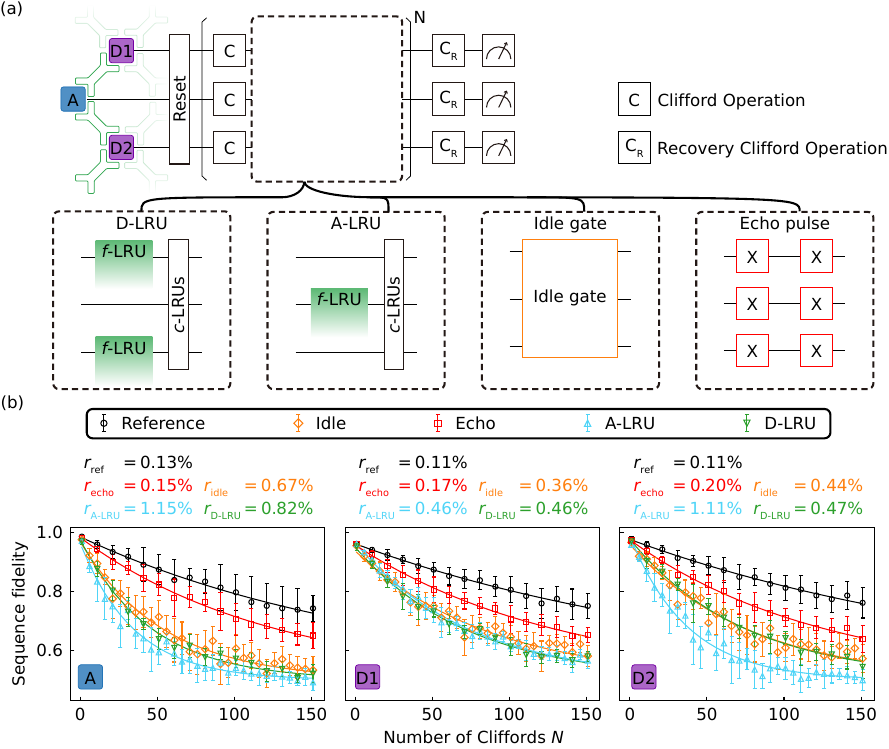}
    \caption{\label{sfig_LRU_error_analysis} 
    Error budget for $f$-LRUs and $c$-LRUs.
    (a) Circuit diagram for the three-qubit RB. The interleaved operation are D-LRU, A-LRU, idle gate and echo operation with two X gates as illustrated. Note that all these operations maintain the same total duration.
    (b) Measured RB sequence fidelity, indicated by the ground state population, versus the number of Clifford gates. 
    The left, middle, and right panels correspond to the results for qubits A, D1, and D2, respectively.
    }
\end{figure}

To evaluate the impact of parallel data $f$-LRUs and the simultaneous $c$-LRUs on the computational subspace, we compare the results of interleaved randomized benchmarking (RB) to those of idle gates over the same duration, as well as a dynamical decoupling (DD) sequence (two X pulses) with the same total duration, as depicted in Fig.~\ref{sfig_LRU_error_analysis} (a). 
The average error ($r_{\rm idle}$) induced by the idle gates is measured to be 0.49\%, which accounts for most of the observed average error (0.58\%) associated with the LRU operations. We observe that the average error ($r_{\rm echo} = 0.17\%$) due to the DD sequence of the same duration is significantly lower, indicating that dephasing or detuning errors are the primary sources of errors in the LRUs or idle gates.

We also benchmark the ancilla $f$-LRU using a similar approach, see Fig.~\ref{sfig_LRU_error_analysis}. We denote the ancilla $f$-LRU and the data $f$-LRU as A-LRU and D-LRU in Fig.~\ref{sfig_LRU_error_analysis} to distinguish these two separate experiments.
Larger errors ($r_{\rm A-LRU} \approx 1.1\%$), are observed on qubits A and D2. This is primarily attributed to the conditional phase accumulation that occurs when the coupler is adiabatically tuned close to the qubits to expedite the parametric interaction.
Notice that the $|e\rangle \rightarrow |f\rangle$ transition frequency of D2 is near resonant with the $|g\rangle \rightarrow |e\rangle$ transition frequency of A. Consequently, there is a relatively strong ZZ interaction in the range of 0.5-1.0 MHz when the coupler frequency is tuned close to the active regime.
This ZZ interaction can be mitigated by altering the coupler's interaction frequency or by employing an additional conditional phase gate for compensation.
The remaining  errors may arise from non-adiabatic transitions during the rising and falling edges of the adiabatic modulation, a mechanism akin to those seen in coupler-assisted conditional phase gates~\cite{xu2020high,chu2021coupler}.


\section{Simultaneous data and ancilla qubit LRUs}

\begin{figure}[H]
    \centering
    \includegraphics{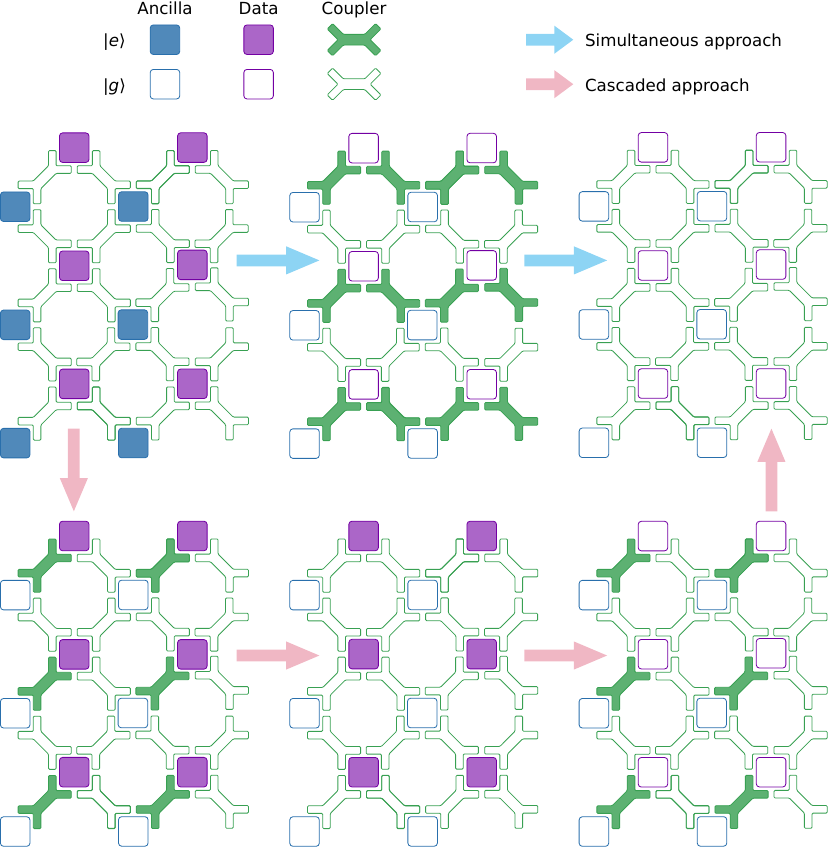}
    \caption{\label{sfig_simul_lru} LRU scheme for data and ancilla qubits in a 2D grid. The blue arrow represents the cascaded data qubit (blue) LRU and ancilla qubit (purple) LRU. The pink arrow represents the simultaneous LRU scheme for data and ancilla qubits using independent couplers. The filled (unfilled) patterns indicates the presence (absence) of leakage population on the qubits (squares) or couplers (bone shape).
    }
\end{figure}

In the main text, LRUs for the data and ancilla qubits are not applied in parallel to avoid crosstalk between qubits coupled to the same coupler. Instead a  cascaded scheme for data qubit LRU and ancilla qubit LRU is adopted, where only one of the four couplers coupled to the same qubit is modulated and only one parametric frequency is applied to the same coupler, as shown in the bottom of Fig.~\ref{sfig_simul_lru}.
Applying simultaneous LRUs for the data and ancilla qubits is possible, although the calibration process is more complicated. 
In a 2D grid, the number of couplers is roughly twice the number of qubits. This abundance of couplers ensures that there is enough capacity to facilitate simultaneous qubit LRUs, as illustrated in the top of Fig.~\ref{sfig_simul_lru}.
However, simultaneously tuning two adjacent couplers that are connected to the same qubit would require detailed calibration of the phase accumulation within the computational subspace and the prevention of unwanted couplings.

\section{Leakage and error syndrome in QEC}

\subsection{Leakage accumulation from high-power readout} 
\begin{figure}[H]
    \centering
    \includegraphics{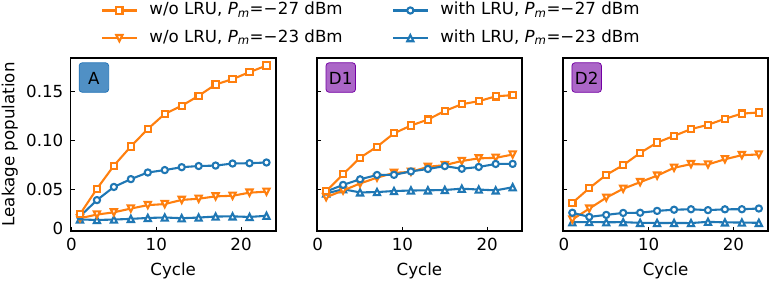}
    \caption{\label{sfig_leakage} 
    Leakage population on each qubit versus parity check cycles, with (blue) and without (orange) $f$-LRUs.
    The dash lines correspond to the data from Fig.4(b) in the main text, corresponding to a lower ancilla measurement power of $P_m=-27$~dBm. 
    An increased leakage population is observed at higher ancilla measurement power ($P_m=-23$~dBm).
    }
\end{figure}

In Fig.~\ref{sfig_leakage}, we show that elevating the measurement power $P_m$ at the electronics output port for the ancilla qubit leads to a substantial increase in leakage accumulation, particularly on the ancilla qubit. We attribute this outcome to readout-induced state leakages~\cite{sank2016measurement}.
It is important to note that the use of $f$-LRUs does not entirely suppress the leakage accumulation, which may be due to the excitation of higher states (such as $|h\rangle$) by the more intense readout power.

\subsection{Intercept leakage error at the 0-th cycle of the stabilizer measurement circuit}

\begin{figure}
    \centering
    \includegraphics{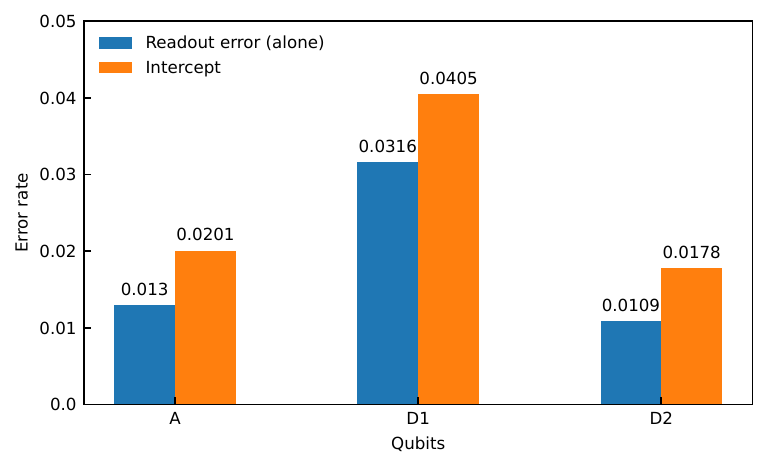}
    \caption{\label{fig_intercept}
    The measured $|f\rangle$ population at the 0-th cycle of the stabilizer measurement circuit (orange), as well as the average measured $|f\rangle$ population when the qubit is individually prepared in $|g\rangle$ and $|e\rangle$ states (blue), for qubit A (left), D1 (middle), and D2 (right). This data was obtained using three-state readout.
    }
\end{figure}

As mentioned in the main text, the nonzero intercept observed at 0-th cycle in Figure 4(b) is mainly caused by readout error. 
We compare the intercept with the single-qubit three-state readout error for all three qubits, as shown in Fig.~\ref{fig_intercept}, and find that the readout error accounts for most of the intercept at the 0-th cycle.
The slight increase in intercept ($\sim 0.007$) compared to the single-qubit readout error is likely caused by crosstalk during the simultaneous readout of three qubits.
    

\subsection{Improvement of QEC from the coupler-assisted LRUs}
\begin{figure}[H]
    \centering
    \includegraphics{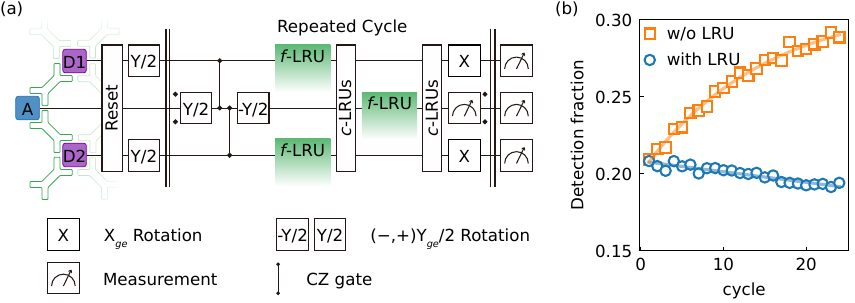}
    \caption{\label{fig_improvement} The error-detection probability versus cycles in weight-2 Z stabilizer measurement. The blue dots / orange squares denote the experiment data with/without LRUs.}
\end{figure}

    In Fig.4(c) of the main text, artificial leakage was injected during the QEC cycles.
    For a more direct comparison, here we show the data of weight-2 Z stabilizer without artificial $|f\rangle$ state leakage injection, as displayed in Fig.~\ref{fig_improvement}. Without applying the $f$-LRUs, the detection fraction of the leakage event increases by 0.08 after 25 cycles. 
    In sharp contrast, with the application of $f$-LRUs, the error detection fraction even decreases slightly as the number of stabilizer cycles increases.

\subsection{Error syndromes with leakage injection at odd/even cycles} 
\begin{figure}
    \centering
    \includegraphics{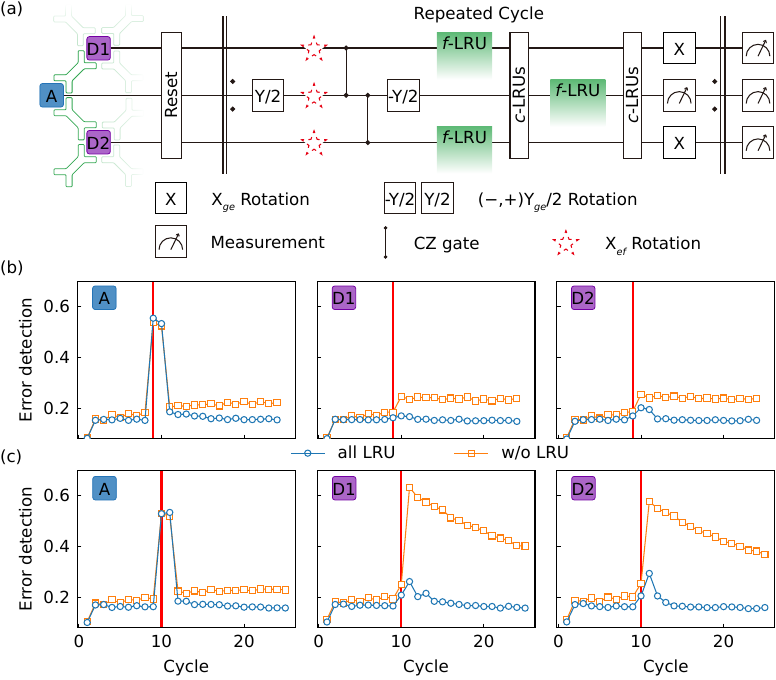}[H]
    \caption{\label{sfig_error_synd}
    Error syndromes resulting from leakage injection at odd and even cycles.
    (a) The Z stabilizer measurement circuit with leakage injections (red dashed stars). 
    The X pulses on data qubits are applied to mitigate phase errors caused the AC-stark shift during ancilla measurement.
    (b) Error syndromes for the 9-th cycle leakage injection on the A (left), $\rm D_1$ (middle), $\rm D_2$ (right) qubits, respectively. The vertical red line indicates the position of injection.
    (c) Error syndromes for the 10-th cycle leakage injection.
    }
\end{figure}

It is worth to clarify that leakage errors injected into data qubits during odd and even cycles show distinct error syndromes. As illustrated in Fig.~\ref{sfig_error_synd}, the error syndrome from leakage injection in the odd (9-th) cycle is significantly less pronounced compared to that in the even cycle just next to it.
This discrepancy is attributed to the echo pulses applied to the data qubits during the ancilla measurement. The logical state, preserved by the stabilizer circuit, oscillates periodically between $|0_L\rangle$ = $|00\rangle$ and $|1_L\rangle$ = $|11\rangle$. In odd cycles, the data qubits are more likely to be in the ground state, which results in a reduced impact from the leakage error induced by the ${\rm X}_{ef}$ rotation.

\bibliography{Reference}